\documentstyle[aps,eqsecnum]{revtex}
\begin{document}
\title{\hspace{12.0cm} Preprint WSU-NP-96-11 \protect\\
\vskip 1.5cm
The wedge form of relativistic dynamics}                            
\author{A. Makhlin}
\address{Department of Physics and Astronomy, Wayne State University, 
Detroit, MI 48202}
\date{August 6, 1996}
\maketitle
\begin{abstract}
It is commonly accepted that in hadronic or nuclear collisions at extremely
high energies the shortest scales are explored.  At the classical level,
this property of the interaction is closely related to the Lorentz
contraction of the fields of colliding particles which provides
instantaneous switching the interaction on.  I argue that the underlying
quantum dynamics should be confined to within the light wedge of the
two-dimensional plane where the first interaction takes place
and suggest to include this property as the boundary condition for the
quantum field theory which describes the collision process.
Connection between the type of inclusive process and the temporal order
of its dynamical evolution is discussed. The one-particle states and
propagators of the perturbation theory for the scalar and fermion fields
are found. 
\end{abstract}                                                           

\section{Introduction}\label{sec:SNI}    

In the previous paper \cite{QGD},  I suggested to  view hadronic and  
nuclear collisions at extremely high energies in a new way; a strongly
localized interaction performs a kind of spectral analysis of the initial
bound states in terms of free quarks and gluons of the final one. 
Following this idea I re-formulated  theory of the deep inelastic electron
proton scattering (DIS) in the language of the quantum field kinetic
theory \cite{QFK} which escapes an intermediate parton phenomenology
obviously used in calculation of the hadronic  processes at high energies. 
The general scheme to calculate the quark and gluon distributions after
the first hard collisions of heavy ions  was traced out in paper
\cite{QGD}. I argued that a collision of two nuclei which are
Lorentz-contracted into two plane sheets creates an environment for the
perturbative regime of interaction between quarks and gluons, which was
impossible before the collision: the perturbative vacuum itself is a
product of the collision. Quantum dynamics of quarks and gluons before and
after collision of the nuclei are qualitatively different. However,
several important issues  were left aside in the previous papers. 

Quantum field theory has a strict definition of the {\it dynamics}. This
notion was introduced by Dirac  \cite{Dirac}  at the end of 40-th in
connection with his attempt to build a quantum theory of gravitational
field. Every (Hamiltonian) dynamics includes its specific definition of
the quantum mechanical observables on the (arbitrary) space-like surfaces,
as well as the means to describe evolution of the observables  from the
``earlier'' space-like surface to the ``later'' one. The ``infinite
momentum frame'' (IMF) is a synonym of the ``light-front dynamics,'' which
had been introduced  by Dirac  as an example of dynamics , along with the
other forms of dynamics.    Besides the light-front dynamics, Dirac has 
also suggested the so-called point form of field dynamics which  was
conceived  as a tool to describe the interaction of the field with the
point-like classical  particle.    

Usually both nuclei before the collision are considered separately,  in
their own IMF, and we have two  Hamiltonian dynamics with different
definition of the time variable.  This is a severe shortcoming of the
theory.  Factorization scheme suggests a detour.
Instead of solving the problem, it replaces the true bound states of the
quark and gluon fields in hadrons  by the artificial flux of free partons.
Factorization strictly requires a scale which is given by some  {\em
measured hard probe},  like high momentum transfer in DIS or the heavy
dilepton mass in the Drell-Yan process. The method is not expected to work
when the experiment does not suggest any ``hard probe.''  Then the idea of
factorization loses its footing and the formal factorization scale becomes
an ill-defined infra-red  cut-off. The heavy-ion collision is just the
case. In fact, the nuclei probe each other at all scales and we are
compelled to consider interaction of the two bounded systems without
appealing to the parton picture \cite{QGD}. Thus, it is imperative to find
a  way to describe quarks and gluons from both nuclei  as well as the
products of their interaction using {\it the same Hamiltonian dynamics}. 
This requirement follows solely from the fact that the precise definition
of the field states (particles) depends on how the global observables are
defined.  

The proper definition of states is highly  nontrivial and important
question because the search of the QGP in heavy ion collision is, in the
first place, the search for the evidence of the entropy production. 
Indeed, before the collision the quark and gluon fields are assembled into 
two coherent wave packets, the nuclei, and, therefore, the initial entropy
equals zero. The coherence is lost and the entropy is created due to
interaction. Though one may wish to rely on the invariant formula, 
$S={\rm Sp}\rho\ln\rho$,  which expresses the entropy $S$ via the density 
matrix $\rho$, at least one basis of states should be found explicitly. 

Amongst the all known solutions of physical problems, those which  were
starting with a lucky guess about the normal coordinates (degrees of
freedom), are most elegant and successful. In QCD (and even in QED),  an
appropriate choice for the gluons (photons) is always difficult because
the  gauge is one of the elements of the Hamiltonian dynamics. The gauge,
as the dynamics itself, is a global object, and both nuclei should be
described using the same  gauge condition.   The light-front form of
dynamics and the light-front gauge proved to be sensible tools to study
the DIS process,  the proton interaction with the  structureless electron,
basically because the gauge is physical. In this gauge, the interference
of the final state gluons is suppressed and the ladder diagrams give the
leading contribution to the DIS cross section.

The primary choice of the degrees of freedom  is successful if,
even without any interaction,  dynamics of the normal modes adequately 
reflects main physical features of the phenomenon. Hence, the normal modes 
of the fields participating in the collision of two nuclei should be
compatible with their Lorentz contraction. Unlike the incoming plane waves
of the standard scattering theory, the nuclei have well defined shape and
the space-time domain of their intersection is well defined. Of ten
symmetries of the Poincare group, survive only  rotation around the 
collision axis, boost along it,  and the translations and boosts in the
transverse directions.   The idea of the  two plane sheets collision
immediately leads us to  the {\em wedge form};  all possible states of
quark and gluon fields before  and after collision must be confined to
within the past and the future  light cones (wedges) with  the $(x,y)$
collision plane as the edge.  It is profitable to limit in advance the set
of normal modes  to those which have the symmetry of the localized
interaction. In this {\it ad hoc} approach, all the spectral components of
the nuclear  wave functions ought to collapse in the  two-dimensional
plane of interaction,  even if all the confining interactions of quarks
and gluons in the hadrons  and the coherence of the hadronic wave
functions are neglected. 

In the wedge form of dynamics, the states of  free quark and gluon
fields are defined (normalized) on the space-like hypersurfaces of the
constant proper time $\tau$, $\tau^2=t^2-z^2$. The main idea of the
approach is to study dynamical evolution of the interacting fields along
the Hamiltonian time $\tau$. The gauge of the gluon field is fixed by the
condition $A^\tau =0$.  This simple idea solves several problems. On the
one hand,   it becomes possible to treat two different light-front
dynamics  which describe  each nucleus of the initial state separately, as 
two limits of this single dynamics.  On the other hand, after collision
this  gauge simulates a  local temporal axial gauge. This feature
provides a smooth transition to the boost-invariant regime of the 
created matter expansion (as the first approximation). 

The feature of the states to collapse in the interaction vertex will be
crucial for understanding the dynamics of the collision. A simplest
optical prototype of the wedge dynamics is  {\em camera obscura}, a
dark chamber with the pin-hole in the wall. Amongst many possible {\em a
priori} ways to decompose the incoming light, the camera selects  one.
Only centered at the pin-hole spherical harmonics can penetrate inside the
camera.  The spherical waves reveal the angular dependence  at some
distance from the center and build up the image on the opposite wall as a
screen. 
            
The dynamical example is  reflection of light from the plane boundary
between vacuum and the media. For monochromatic signal the solution of the
problem is given by the Fresnel formulae which connect the amplitudes of
the incoming, reflected and refracted waves. These triplets of waves can
be viewed as the stationary states of the photons  which obey the boundary
condition. The latter  depend on the frequency of the photon (due to
dispersion of the refraction index) and on coordinate of the boundary (let
it be the plane $z=0$).  Consider now the non-stationary process, when the
electromagnetic wave has a front, $z-t=0$, and is not monochromatic.  The
problem is linear and can be easily solved by presenting the light pulse
as a superposition of the stationary solutions. Every stationary partial
wave triplet  knows nothing about its origin from the pulse. However,
before the light front has reached the boundary  $z=0$, the reflected and
refracted components of the triplets add up in such a way, that the
refracted and reflected fields do not exist along with the field of the
incoming wave ahead of the leading wave front. Mathematically, this is
possible due to the proper balance of the amplitudes and phases in
spectral decomposition of the signal. The first study of the wave front
propagation in dispersive media was done by Sommerfeld and Brillouin at
the beginning of the century \cite{precurs} and had the goal to show that
even in the case of anomalous dispersion, when the group velocity may
exceed the velocity of light, the principles of special relativity are not
violated. The full study of the  transient process on the boundary of the
media was done in Ref.~\cite{SMKS}.

The problem of the light front in the medium can be re-formulated in terms
of quantum mechanics  as well. The system of wave triplets may be
quantized as any other system of the field oscillators, {\em etc.}  Then
emission of any new wave is just excitation of one of these oscillators or
superposition of such excitations.  There are infinitely many sets of 
wave triplets which correspond to the different positions of the boundary,
$z=a>0$, which may be used for spectral decomposition of the incoming wave
at $z<0$.  The {\em incoming photon} can be considered as superposition of
triplets from any of these sets since all these superpositions are virtual
until the  interaction of the wave with the media  selects one of them as
a consequence of the real process.  Until $~a~$ is an arbitrary
parameter, these sets respect the translational symmetry of the free space
in $z$-direction.  Interaction of the wave front with the medium makes an
explicit selection, ~$a=0$, and destroys the initial symmetry. This simple
example illustrates general principle: of many possible decompositions of
the initial state, real is the one,  corresponding to actual interaction.
The  symmetry of the system, as it is seen  before and after interaction,
may be different.

Our further concern is localization of the interaction in collisions of
the particles. This issue includes different questions. We say that
electron, muon or quark are point-like since the lowest order vertex of
their electromagnetic interaction does not require form-factor. However,
this reflects only locality of the interaction between the fundamental
fields of QED.  Isolation of the domain in space and time where this local
interaction happens is a quite separate question. Previous example tells
us that the conclusive judgement about localization of interaction in the
collision event can be made only {\em post factum}. Indeed, colliding two
protons at very high energy, one cannot anticipate, how many particles will 
be produced in the final state. On average, the multiplicity is quite low,
since the total cross section is always dominated by the soft peripheral
processes.  Only small fraction of events will have  high multiplicity.
This narrow  subset consists of those were the geometry of interaction
(transition currents) is effective enough with respect to  excitation of a
special  set of modes.  It is direct experimental evidence that at highest
energies and highest multiplicities the spectrum of hadrons tends to have
a plateau in the central rapidity region. Extrapolating this result to
infinite energy, one immediately concludes that the clue to understanding
this process is the Lorentz contraction of the colliding  particles
\cite{Landau}. To produce the wide rapidity plateau, the system should
lose any initial geometrical scale along with translational invariance in 
time direction and direction of the collision axis.  The wedge form of
dynamics has this  {\em kinematic} property in its initial formulation and
employs it as the classical boundary condition  for the underlying
dynamics of quantum fields. Phenomena with this kinematic  regime can be
met both in QED and QCD. However, only in QCD can we encounter the
necessary ratio of scales:  the typical scale of the vacuum fluctuations
which leads to the confinement ({\em e.g.} size of the instanton) is of
the order 0.3{\em fm}, while already at 200{\em AGeV} nuclear collision
the resolved longitudinal scale is 0.1{\em fm}. Thus, there is a  deep
reason to rely  on the wedge form of dynamics in most of the processes
where the gluonic  degrees of freedom are actively excited. A part of this
excitation is creation of the new ground state, {\em the  perturbative
vacuum}. This state exists since the moment of first overlapping of the
Lorentz-contracted hadrons and until hadronization  completely washes it 
out at the proper time $\tau_{hadr}$ .  Only after the  hadronization
process ends up, the system  recovers its {\em translational invariance on
hadronic scale}, and, if the rapidity plateau was infinite (in the limit
of the infinite c.m.s energy), it never would. In fact, the intermediate
perturbative dynamics of quark and gluon fields always takes place in the
regime of the broken translational invariance!

By its logic, the wedge dynamics is contiguous both to the standard
quantum   scattering theory which relies on the parton picture and deals
with the plane wave asymptotic states, and to the theory suggested by
McLerran and Venugopalan \cite{Raju,Raju2,Larry}, which starts with the
picture of classical evolution of the gluon fields of colliding quark
sources and treats quantum effects as small corrections to the classical
background. However, there are several major differences. 

(1) ~I do not view quarks as the sources of the gluon fields in hadrons or
even in nuclei. Rather, I keep in mind the picture which is based on the
long  standing observation by Shuryak \cite{Shuryak1} that the energy
scale of confinement is weak, while the energy density of QCD vacuum is
very high. In his  approach, the hadrons are thought of as the
point-to-point correlators of various quark currents in  physical vacuum
and  kept localized due to the tunneling.  Recent development of this idea
proved to be very successful in computation various properties of hadrons
\cite{Shuryak2}.  Though connection between the Euclidian regime of these
calculations and  Minkowskian dynamics of the real world is poorly
understood, the above mentioned ratio of energy scales should be 
considered the most  significant physical observation. In the advocated
approach, the hadrons in the QCD vacuum can be compared to the sound waves
in a heavy fluid; though the energy and momentum of the acoustic pulse may
be very high, the atoms of the fluid experience very small displacements.
However, nonlinear interaction of two waves may cause cavitation. To cause
cavitation of the fluid by sound waves, one needs high  gradients of the 
acoustic displacements (high tension) rater than high amplitude of the
acoustic wave.  I suggest to view  the multiparticle production in
hadronic collision like a signature of a similar phenomenon;
``cavitation'' of the physical QCD vacuum and creation of the domain with
the perturbative dynamics of quarks and gluons. To switch it on, one needs
extreme Lorentz contraction which resolves fluctuations of much shorter
scale than those providing a natural  ``soft'' confinement in the hadrons. 

(2) ~Unlike in Refs.~\cite{Raju,Raju2,Larry}, I do not require any
specific information about the wave functions of the hadrons or nuclei.
Such a knowledge is needed to compute the hadronic or nuclear
form-factors.  In deeply inelastic processes  one needs wave function only
to support the concept of the finite size object and its Lorentz
contraction. The structure functions of the deep inelastic processes
reflect properties of the physical vacuum rather than of the hadrons.

(3) ~I view any deeply inelastic process as a transient process
developing in space and time. I am interested primarily in  the first
few fermi of its history, before the asymptotic states of scattering are
formed. During this period, longitudinal and  transverse fields are not
separated geometrically, as they never are in the IMF. In parallel studies
I show that the spurious poles of the null-plane gauge carry indication
of static configuration of the gauge field \cite{TAG}. Analysis of the
evolution equations for the observables of the inclusive process carried
out in Ref.~\cite{HQ} shows that  longitudinal and the transverse fields
are not separated dynamically. Moreover, examination of the extended set
of evolution equations for the structure functions of DIS brings to light
a new element, the feed-back via longitudinal fields, which 
is consistent only with the power-like enhancement of
the DIS cross section at low $x$.  In the wedge dynamics, which
incorporates geometry of the DIS process already in the structure of
states of free fields, longitudinal and  transverse fields are not
separated even mathematically \cite{WDG}.

(4) ~Analysis of Ref.~\cite{HQ} explicitly shows that  an attempt to
extend the IMF technique beyond the utilitarian needs of the parton model
brings in severe singularities which are  intrinsic for the IMF dynamics.
Their physical regularization undermine status of $x_F$ as physical
variable. The wedge dynamics naturally cures this problem in such a way
that  {\em geometry of interaction} between the fields is {\em not} 
contracted  into the  singular geometry of plane sheets.

Since the wedge form of dynamics relies heavily on the localization of the
interaction that physically resolve the QCD degrees of freedom, certain
pre-requirements should be met.  The idea of the point-like  localization
of the interaction is unambiguous for the finite-size objects like nuclei
or  hadrons in the environment of the process which resolve their internal
structure.   In the system of two  nuclei, localization  is, perhaps,
perfect, and all dynamics takes place in the future wedge of the
intersection plane. 

Geometrical similarity of the $ep$-- and $pp$--collisions is the major
base for hope that  the $ep$-DIS data contain the same information about
the process of the 
 proton destructure which is essential in the $pp$- and $AA$-collisions.  The
deeply inelastic interaction between the structureless electron and the
hadron carries all necessary signs of localization which allow one to
treat this process within the boundary conditions of the wedge dynamics.
However,  since QED has no intrinsic scale similar to 
hadronic scale $\Lambda_{QCD}$ of quantum chromodynamics,
one should be careful with an {\em a priori} localization of the
structureless electron on the QCD scale, since this step would mean an
assignment of the structure function to the electron itself!  
For this reason, I begin  in section~
\ref{sec:SN1} with the discussion of how the wedge dynamics solves the
problem of localization of the interaction. In section \ref{sec:SN2}, ~I
discuss how the trigger in the measurement of the DIS cross section may
affect the temporal order of the processes that takes place near the light
cone.  In Sec.~\ref{sec:SN3} and  Sec.~\ref{sec:SN4} I present 
Hamiltonian formalism for the scalar and  the fermion fields in  the wedge
form of dynamics. The one-particle solutions for the free fields and their
Wightman functions and propagators are obtained. These are necessary for
the future formulation of the perturbation theory. Even for the free gauge
field, calculation of the gluon  modes and the gluon propagator is a
complicated mathematical  problem. Solution of this problem is described in
the separate paper \cite{WDG}.                                                  

\section{Localization of interactions in wedge dynamics}
\label{sec:SN1}    

For the further qualitative  analysis of the approach it is enough to
use the one-particle wave functions (\ref{eq:E2.11}) of the scalar
field. Let us write down  the wave function $\Xi_{\theta,p_\perp}(x)$ in the
following form:
\begin{eqnarray}
\Xi_{\theta,p_\perp}(x)= {1 \over 4\pi^{3/2}}  
e^{-im_{\perp}\tau\cosh(\eta-\theta)}
e^{i{\vec p}_{\perp}{\vec r}_{\perp}}~~. 
\label{eq:EI.1}\end{eqnarray}      
This function describes a state of the scalar particle  which occupies the
future and the past light wedges of the collision plane. The above form
implies that $\tau$ is positive in the future of the wedge vertex  and
negative in its past and, as usually, $m_{\perp}^{2}=p_{\perp}^{2}+m^2$. 
Though this wave function is an
obvious plane wave, it carries the  quantum number $\theta$, rapidity of
the particle, instead of the momentum $p_z$. The main physical difference
is that the wave function   (\ref{eq:EI.1}) is normalized on the
hypersurface $\tau=const$. At large $~m_{\perp}|\tau|$, the phase of the
wave function  $\Xi_{\theta,p_\perp}$ is stationary in a very narrow
interval  around $~\eta=\theta~$ (outside this interval, the function
reveals oscillations with exponentially increasing  frequency); the wave
function describes a particle with rapidity $~\theta$ moving along the
classical trajectory .  However, for $m_{\perp}|\tau|\ll 1$, the phase 
is almost constant along the surface $\tau=const$~.  The
smaller $\tau$, the more uniformly the domain of  stationary  phase is
stretched along the light cone. A single particle with the wave function
$\Xi_{\theta,p_\perp}$ begins its life as the wave  with the given
rapidity $\theta$ at large negative $\tau$. Later, it becomes spread over the
boundary of the past light wedge at $\tau\to -0$. Being spread, it
appears on the boundary of the future light wedge at $\tau\to +0$.
Eventually, it again becomes a wave  with  rapidity $\theta$ at large
positive $\tau$.  The size and location of the interval where the phase
of the wave function is stationary will play a central role in all
subsequent discussion, since it is equivalent to the  localization.
Indeed, overlapping of the domains of stationary phase in space and
time provides the most  effective interaction of the fields.

The size $\Delta\eta$ of the $\eta$-interval  around the particle rapidity
$\theta$, where the wave function is  stationary, is easily evaluated.
Extracting from the exponential of Eq.~(\ref{eq:EI.1}) the trivial factor
$e^{-im_{\perp}\tau}$ which defines evolution of the wave  function in the
$\tau$-direction, we obtain an estimate:
\begin{eqnarray}
 2~m_{\perp}\tau\sinh^2(\Delta\eta/2)\sim 1~~. 
\label{eq:EI.2}\end{eqnarray}      
The two limit cases are as follows, 
 \begin{eqnarray}
 \Delta\eta \sim \sqrt{2\over m_{\perp}\tau},
~~{\rm at}~~ m_{\perp}\tau \gg 1~,~~~{\rm and}~~~
  \Delta\eta \sim \ln {2\over m_{\perp}\tau},
~~{\rm at}~~ m_{\perp}\tau\ll 1~~.
\label{eq:EI.3}\end{eqnarray}   
In the first case one may boost this interval into laboratory reference
frame and see that the interval of  stationary phase is Lorentz 
contracted (according to the rapidity $\theta$) in $z$-direction. 

Obviously, one may wish to deal with the wave packets built from the waves
with different transverse momenta.  The packets of the waves with the same
rapidity $\theta$ do not experience dispersion in the longitudinal
direction. [The language of the rapidity is extremely useful here since it
supports an intuitive picture of the partial waves as co-movers which
constitute the proton.] This kind of expansion can be employed to form the
finite size objects. The amplitudes and the phases of the coefficients in
this expansion are balanced in such a  way that before the collision the
wave packets represent the well shaped nuclei.  If two localized objects
simultaneously pass through the vertex, then the partial waves that form
their wave functions effectively overlap in the vicinity of the light
wedge. The interval of time,  when the partial wave with transverse
momentum $p_t$ is spread, is of the order $\tau\sim1/m_t$. The high-$p_t$
components of the wave function assemble around the world line of the
initial rapidity $\eta=\theta$ earlier, and have less time to interact 
than the low-$p_t$ components. Perhaps, it is a good approximation to view
the proton after it has passed the interaction vertex as a superposition
of various plane waves with the same rapidity $\theta$. If no interaction
with the electron occurs at sufficiently small $\tau$, then the amplitudes
and the phases of these waves remain unchanged and the packet assembles
into the initial proton.  If only small fraction of partial waves with not
too high $p_t$ has interacted at not too small $\tau$, then the entire 
process may become diffractive: the high-$p_t$ components of the wave
packet will assemble quite early along the proton's world line and build
up a core of a slightly disturbed proton.

In general, Lorentz-contraction is a feature of the finite size objects. 
However, it is well known that quantum cross sections of the Born's
approximation  coincide with those obtained classically.  Classical
treatment explicitly accounts for the Lorentz transformation of the fields
of the colliding particles. It alone leads to  the localization of the
interaction and builds up a geometrical picture of the plane sheets. 
Quantum mechanics  deals with interaction of the waves and does not
support the image of the Lorentz contracted classical field. Therefore,
the physical localization of the interaction in the wave dynamics requires 
localization of the transition currents which is provided by the wave
functions. 

Let the particle has rapidities $\theta$ and $\theta'$ and transverse
masses $m_{\perp}$ and $m'_{\perp}$ before and after scattering,
respectively.  The change of its state is due to the transition current,
\begin{eqnarray} 
j_{\bf k, k'}(x_1) \sim e^{-i\tau[m_{\perp}-m'_{\perp}]} 
e^{-i\tau[m_{\perp}\sinh^2{\theta-\eta_1 \over 2}
-m'_{\perp}\sinh^2{\theta'-\eta_1 \over 2} ]}~~.
\label{eq:EI.4}\end{eqnarray}
  
For the process with high transverse momentum transfer  we have 
$~m'_{\perp}\gg m_{\perp}$.  It is easy to see that at large difference
$\Delta\theta=\theta -\theta'$  the stationary phase intervals of the two
modes overlap only at sufficiently  small  $\tau$, when
$m_{\perp}\tau\sim e^{-\Delta\theta /2}$.   At this moment, the long
stationary phase domain of the incoming particle ($m_{\perp}=m$ is low)
begins to cover still narrow stationary phase domain of the scattered
particle ($m'_{\perp}$ is high). Similar analysis is valid for the
transition current of the second particle, $j_{\bf p, p'}(x_2)$. Distance
between the points $x_1$ and $x_2$ along the surface $\tau =const$ is  
$R=[({\vec r}_{\perp 1}-{\vec r}_{\perp 2})^2+ 
\tau^2\sinh^2(\eta_1-\eta_2)]^{1/2}$.  The Coulomb part of the vector
field propagator in the gauge $A^\tau=0$ is proportional to $ R^{-2}$
\cite{WDG}. At very small
$\tau$, even at finite rapidity difference  $\Delta\theta$, the distance
between transition currents  $j_{\bf k, k'}(x_1)$ and $j_{\bf p, p'}(x_2)$
is entirely due to their  transverse separation,  
$|\vec{r}_{\perp 1}-\vec{r}_{\perp 2}|$. Therefore, the process of
scattering indeed takes place in the vicinity of the wedge vertex, in
compliance with the classical treatment. Moreover, interaction that excites 
mode with  rapidity $\theta'$ and transverse momentum $k'_{\perp}$
takes place at small $\tau\sim 1/k'_{\perp}$  at some interval
of $\eta$ around $\theta'$.  It virtue of (\ref{eq:EI.3}), the higher 
$k'_{\perp}$, the narrower this
interval is. By the current conservation, both initial- and final-state
wave functions should effectively overlap on this interval.

The last conclusion is in strong correlation with an observation
stimulated by the study of the gluon correlator in the gauge $A^\tau=0$ 
\cite{WDG}. The limit of  the gluon propagator of this gauge near the
null-plane $x^-=0$ is the propagator of the gauge $A^-=0$. Close to the 
plane $x^+=0$ the same propagator corresponds to the gauge $A^+=0$. The
gauge $A^+=0$ is obviously used to describe the QCD evolution of the
proton with momentum $P^+\rightarrow\infty$ and $P^-=0$ which enters into
collision with the electron along the null-plane $x^-=0$.   However, the
gluon propagator of the gauge $A^+=0$ is found near the opposite boundary
of the light wedge, where, according to the wedge dynamics, the
electromagnetic transition current extinguishes the old  wave function of
the quark which was ``prepared'' for the last interaction  by the QCD
evolution, and excites the new one, with the highest $p_{\perp}^2\sim Q^2$.
This is possible only if at least the latest stage of the QCD evolution
takes place near the null-plane $x^+=0$! Thus, the  existing theory of QCD
evolution seems to support  the geometry  of high-energy collision
suggested by the wedge form of dynamics.

What happens at other rapidities, where the transition currents are not
stationary? Though the currents experience rapid oscillations in 
$\eta$-direction, and conditions for the high-$p_t$ jet formation are not
fulfilled, the wave functions overlap there as well. Being unable to
produce the coherent jet, these fragments of the wave functions should
break up into many soft quanta which, either uniformly or with gaps,
fill in  the rapidity interval between the leading jets.  

We arrived at the major conclusion that at high relative rapidity of the
colliding particles the most effective interaction takes place in the
vicinity of the light cone.  This conclusion is very close to the
observation made by Ioffe long ago, that the lepto-production process is
dominated by the close vicinity of the light cone and long  distances in 
the direction of the
virtual photon propagation \cite{Ioffe}.   However there is conceptual
difference between the Ioffe's and the present motivations of this picture. 

The first step of Ioffe's approach is to use the translation invariance
and to replace  the product $j^\mu(x)j^\nu(y)$ of two hadronic currents in
the matrix element of the inclusive process by their  commutator
$[j^\mu(x),j^\nu(y)]$. By causality, the latter vanishes at space-like
separation between $x$ and $y$.  The second step is to employ the Bjorken
scaling as a pre-requirement.

The first of these steps is somewhat ambiguous since the momentum transfer
$q^\mu$  in the lepto-production is space-like,  $q^2<0$. Therefore, the
required inequality, $q^0>0$, is not Lorentz invariant and the replacement
of product of the two currents,~ $jj$, by their commutator,~ $[j,j]$, is
possible only in a restricted kinematic region or in a special reference
frame. The proximity to the light cone in the lepto-production process, as
it emerges from this derivation, does not look (unlike the light cone
itself) as a Lorentz-invariant feature. However, the plateau in the
rapidity distribution of hadrons at extreme c.m.s. energies says that it
does. The wedge form of dynamics avoids the reference frame dependent
arguments.  The Ioffe's long distances are replaced by the large
intervals of the rapidity coordinate $\eta$. The entire picture can be
boosted from one reference frame to  another  without any changes in
physical interpretation or qualitative change of the normal modes.


\section{Temporal order in inclusive processes}
\label{sec:SN2}

The program of computing the quark and gluon distributions in heavy ion
collisions relies on the data obtained in seemingly more simple processes,
like $ep$-DIS, {\em etc.} It turns out that differently triggered sets
of data may carry significantly different information. In this section,
we discuss two examples and show that even minor change of the way to
take the data may strongly affect the type of the studied process.

The early discussion of the role of the light cone distances in the high
energy collisions \cite{GIP} resulted in Gribov's idea of the two-step
treatment of the inelastic processes \cite{Gribov}:  the gamma-quantum
{\em first} decays into virtual hadrons and {\em later} these hadrons
interact with the nuclear target. This idea looks very attractive since it 
explains the origin of two leading jets in electro-production events,
corresponding to  the target and the projectile (photon) fragmentation. It
provides reasonable explanation of the plateau in the  rapidity
distribution of the hadrons also. However, this elegant qualitative
picture contains a disturbing element, the way how  the words ``first''
and ``later'' are used. 

The question of temporal sequence in relativistic quantum mechanics is
two-fold. The first issue is trivial; any statement concerning the time
ordering  or causality should respect the light cone boundaries. The
second one concerns the nature of quantum-mechanical evolution which
allows one to read out any dynamical information only after the evolution
is interrupted by the measurement. This feature is usually referred to as
a collapse of the  wave function.  All dynamics of the system takes place
before the measurement. It was already pointed out \cite{QGD} that the
technique of QFK explicitly supports this principle. The Gribov's idea of
the initial fragmentation of the $\gamma^\ast$ in lepto-production process
contradicts it. Indeed, the only quantity measured  in the deep inelastic
scattering experiment is the momentum of the final state electron.
Therefore, the process of the momentum transfer to the electron  must be
the last one in temporal sequence. The off-mass-shell $\gamma$-quantum
which provides this transfer must be a product of the preceeding 
evolution of the hadronic degrees of freedom. Below, we shall show how
this result emerges in the QFK version of the $S$-matrix approach.  

The Gribov's picture will be shown to describe different class of
experiments, where the final state electron is entirely off the control
and the event is triggered by the high $p_t$ quark or gluon jet in the
final state.  In fact, the two leading jets  corresponding to the target
and the projectile fragmentation are present in both types of events,
tagged either by the high $p_t$ electron or by the high $p_t$ jet. 
Regardless the dynamical details, the hadronic distribution is driven by
the geometry of the transition currents near the light wedge of the
interaction plane.  
 
Let us consider  collision process with  parameters  of only one final 
state particle  explicitly measured. Let  this  particle be the electron 
with momentum ${\bf k}'$ and spin $\sigma'$.  The deep inelastic 
electron-proton scattering is an example of such  experiment.  All vectors 
of final states which are accepted into the data ensemble are of the
form  $ a^{\dag}_{\sigma'}({\bf k}')|X\rangle $  where the vectors
$|X\rangle $  form a full set. The initial state consists of the electron
with momentum ${\bf k}$ and spin $\sigma$ and some other particle or
composite system carrying quantum numbers $P$.  Thus the initial state
vector is  $ a^{\dag}_{\sigma}({\bf k})|P\rangle $.  The 
inclusive transition amplitude reads as
$\langle X|a_{\sigma'}({\bf k}')~S~a^{\dag}_{\sigma}({\bf k})|P\rangle $  
and the inclusive momentum distribution of the final state electron is the
sum of the squared moduli of these amplitudes over the full set of the
non-controlled states $|X\rangle$. We obtain the following formula,
\begin{eqnarray}
{ d N_e\over d {\bf k}'}=
\langle P|a_{\sigma}({\bf k}) S^{\dag} a^{\dag}_{\sigma'}({\bf k}')
a_{\sigma'}({\bf k}') S  a^{\dag}_{\sigma}({\bf k}) |P\rangle ~,
\label{eq:E1.1}    
\end{eqnarray}      
which is just an average of the Heisenberg operator of the number of 
the final state electrons  over the initial state.  Since state
$|P\rangle$ contains no electrons, one may commute electron creation and
annihilation operators with the $S$-matrix and its conjugate $S^{\dag}$
pulling the Fock operators $a$ and $a^{\dag}$  to the right and to the
left respectively. Let $\psi_{{\bf k}\sigma}^{(+)}(x)$ be the one-particle 
wave function of the electron. Then the procedure results in
\begin{eqnarray}
{ d N_e\over d {\bf k}'}={1\over 2} \sum_{\sigma\sigma'}
\int dxdx'dydy'{\overline\psi}_{{\bf k}\sigma}^{(+)}(x)
{\overline\psi}_{{\bf k'}\sigma'}^{(+)}(x')
\langle P| {\delta^2 \over \delta{\overline\Psi}(x) \delta \Psi(y)}
\bigg( {\delta S^{\dag} \over \delta\Psi(y')}
{\delta S \over \delta {\overline\Psi}(x')}\bigg) |P\rangle 
\psi_{{\bf k}\sigma}^{(+)}(y)\psi_{{\bf k'}\sigma'}^{(+)}(y')
\label{eq:E1.2}    
\end{eqnarray}      
Introducing the Keldysh convention about the contour ordering
\cite{Keldysh,QFK}, we may rewrite this expression as        
\begin{eqnarray}
{ d N_e \over d {\bf k}'}= {1\over 2} \sum_{\sigma\sigma'}\sum_{AB}
(-1)^{A+B} \int dxdx'dydy'{\overline \psi}_{{\bf k}\sigma}^{(+)}(x)
{\overline \psi}_{{\bf k'}\sigma'}^{(+)}(x')
\langle P| {\delta^4 S_c \over \delta {\overline\Psi}(x_A)
\delta{\overline\Psi}(y'_1)
\delta\Psi(x'_0)\delta\Psi(y_B)} |P\rangle 
\psi_{{\bf k}\sigma}^{(+)}(y)\psi_{{\bf k'}\sigma'}^{(+)}(y') 
\label{eq:E1.3}    
\end{eqnarray}    
where $S_c=S^{\dag}S $.  The electron couples only to the
electromagnetic field. Therefore, to the lowest order,
\begin{eqnarray}
{ d N_e\over d {\bf k}'}={1\over 2} \sum_{\sigma\sigma'}
 \int dxdy{\overline \psi}_{{\bf k'}\sigma'}^{(+)}(y)
{\overline \psi}_{{\bf k}\sigma}^{(+)}(x)
\langle P|\not\!{\bf A}(x) \not\!{\bf A}(y) |P\rangle 
\psi_{{\bf k}\sigma}^{(+)}(y)\psi_{{\bf k'}\sigma'}^{(+)}(x) ~~,
\label{eq:E1.4}    
\end{eqnarray}    
where  ${\bf A}(x)$ is the Heisenberg operator of the electromagnetic
field. Already at this very early stage of calculations, the answer has a
very clear physical interpretation. Once only the final state electron is
measured, the probability of the electron scattering is entirely defined
by the electromagnetic field produced by the rest of the system evolved
from its initial state till the moment of interaction with the electron.
The trigger for this measurements is presence of the electron with high 
enough transverse momentum among the secondaries. 

We may obtain the answer either using equations of QFK 
\cite{QFK,QGD}, or iterating Eq.~(\ref{eq:E1.4}) by means of the
Yang-Feldman equation, $~{\bf A}(x)=\int d^4y D_{ret}(x,y){\bf j}(y)$,~
where $~{\bf j}(y)~$ is the Heisenberg  operator of electromagnetic current
and $~D_{ret}(x,y)~$ is the retarded propagator of the photon. 
Summation over the electrons spins invokes the leptonic tensor 
$~L_{\mu\nu}(k,k')$.~  If $~q=k-k'~$ is the space--like momentum transfer,
then the DIS cross-section is given by
\begin{equation}
 k'_{0} {d\sigma \over d{\bf k'}}= {i\alpha \over(4\pi)^2}
{L_{\mu\nu}(k,k')\over (kP)} 
\Delta_{ret}(q)W^{\mu\nu}(q)\Delta_{adv}(q)~~, 
\label{eq:E1.6}
\end{equation}   
where $W^{\mu\nu}(q)$ is the standard Bjorken notation for the 
correlator of  two electromagnetic currents,
\begin{equation}
 W^{\mu\nu}(q)={2V_{lab}P^{0}\over 4\pi}
\langle P|{\bf j}^\mu(x){\bf j}^\nu(y)|P\rangle~~.
\label{eq:E1.7}\end{equation} 
where $V_{lab}$ and $P^\mu$ are the normalization volume and the momentum 
of the proton in the laboratory frame. Correlator of the currents is the
source of the field which has scattered the electron. Here, both  photon
propagators are  retarded and respect the causal order of the process.   
Since the transition currents  in the interaction of the ultrarelativistic
particles are confined to the nearest vicinity of the light cone, the
simplest geometric arguments support the picture when the last process of
the electron scattering is due to the transition current which acts in the
future of  the wedge vertex.

Any details of the processes that take place in sector of strong
interaction remain unobserved in this type of measurement. The observed
ones are only absorbtion of the electron from the initial state and
excitation  the final state mode of the electron field which take place
with respect to the normal translation-invariant QED vacuum and should
not be localized.  It is the hadronic plateau  which indicates that the
electromagnetic transition current of the strongly interacting quarks was 
localized together with all quark-gluon dynamics. The reason to treat QED
and QCD vacua in so different manner is that of these two,
only QCD has an intrinsic scale, $\Lambda_{QCD}$. 

What happens if we change the observable in the same deep inelastic
process initiated by the $ep$-interaction. Let us  trigger events on the
high-$p_t$ quark or gluon jet in the final state regardless the momentum
of the electron. Then the data ensemble includes the states 
$\alpha_{\sigma'}({\bf p}) |X\rangle$, with the inclusive quark jet, where
$\alpha_{\sigma}({\bf p})$ is the Fock  operator for the final state
quark. The inclusive amplitude for this  process  is   $\langle
X|\alpha_{\sigma'}({\bf p})~S~ a^{\dag}_{\sigma}({\bf k})|P\rangle$.  To
produce the quark,  the hadronic target has  to be hit either by the
photon coming from the first step of virtual  fragmentation of the
electron or by the partons, the electron has previously  fragmented into.
Now we encounter quite different type of fluctuations and both initial
electron and the proton should be treated in the wedge form of dynamics.
Thus, the change of the trigger  drastically affects information read out
of the data. Trigger on the high-$p_t$ jet in the same process allows one 
to filter out fluctuations corresponding to the Gribov's picture. 

This type calculations, extended to the higher orders in coupling
\cite{QGD} have lead to the conclusion  that the QCD evolution describe
(in a real time scale) the gradual process of  excitation a special wave
packet which is ``resonant'' to the interaction in a given inclusive
measurement . Within the wedge form of dynamics one can learn more about
where the process takes place in space and time.

\section{States of scalar particles participating the planar interaction}
\label{sec:SN3}    

From the physical motivation of the previous section it is clear that the
wedge form of dynamics will require special tools for the consistent
development of its mathematical formalism. To begin with, I  shall discuss
the wedge form of the dynamics of the charged scalar particles. They are
not expected to participate in the heavy ion collision at its early stage.
However, it is instructive to work out this simple case since its
 mathematics is very simple and allows one to illustrate the main
ideas without extraneous details.

\subsection{The classical treatment}
\label{subsec:SB21} 

Let us assume that the collision occurs in the plane $t=0,\;z=0$. Two
planes, $~t=z$ and $t=-z$,~ divide the whole space-time into four domains,
future (F), past (P), left (L) and right (R) with respect to the collision
plane.  In these domains we shall use the following coordinates:
\begin{eqnarray}
F:~~t=\tau \cosh \eta,~z=\tau \sinh \eta,~~~~~ 
P:~~t=-\tau \cosh \eta,~z=-\tau \sinh \eta,\nonumber\\
L:~~t=-\tau \sinh \eta,~z=-\tau \cosh \eta,~~~ 
R:~~t=\tau \sinh \eta,~~~z=\tau \cosh \eta~. 
\label{eq:E2.1}    
\end{eqnarray} 
These coordinates induce the metric, different in the different domains,
\begin{eqnarray}
(FP): ds^2=d\tau^2-dx^2-dy^2-\tau^2 d\eta^2,~~ 
(LR): ds^2=\tau^2 d\eta^2-dx^2-dy^2 -d\tau^2~.  
\label{eq:E2.2}    
\end{eqnarray}   
Any normal field theory begins with the action. For the complex scalar field 
the action is
\begin{eqnarray}
{\cal A}=\int d^4 x \sqrt{-g}{\cal L}(x)= 
\int d^4 x \sqrt{-g} [{\rm g}^{\mu\nu}\partial_\mu \phi^*(x)
\partial_\nu \phi(x) - m^2 \phi^*(x)\phi(x)]~~.
\label{eq:E2.3}    
\end{eqnarray}                    
Variation of the action with respect to the fields $\phi$ and $\phi^*$
yields the Lagrangian equations of motion,
\begin{eqnarray}
\partial_\mu [(-{\rm g})^{1/2}{\rm g}^{\mu\nu}(x)\partial_\nu \phi(x)]
+(-{\rm g})^{1/2} m^2\phi(x) = 0,\;\;\; 
\partial_\mu [(-{\rm g})^{1/2}g^{\mu\nu}(x)\partial_\nu \phi^*(x)]
+(-{\rm g})^{1/2} m^2\phi^*(x) = 0 
\label{eq:E2.4}\end{eqnarray}                    
and invoke the locally conserved $U(1)$--current,
\begin{eqnarray}
 J_\mu (x)= i \phi^*(x){\stackrel{\leftrightarrow}{\partial_\mu}}\phi (x),
\;\;\; (-{\rm g})^{-1/2}\partial_\mu [(-{\rm g})^{1/2}
{\rm g}^{\mu\nu}(x)J_\nu (x)]=0
\label{eq:E2.5}\end{eqnarray}          
We shall start constructing the one-particle solutions from the
F-domain, were the final states are supposed to be localized.
Here, the corresponding wave functions of 
the free scalar particles obey the Klein-Gordon equation,       
\begin{eqnarray}
{1\over\tau}{\partial  \over \partial \tau}  
(\tau {\partial \phi \over \partial \tau}) - 
{1\over\tau^2}{\partial^2\phi  \over \partial \eta^2}
-\nabla_{\bot}^{2}\phi +m^2 \phi =0~~,~~~
{1\over\tau}{\partial  \over \partial \tau}  
(\tau {\partial \phi^* \over \partial \tau}) - 
{1\over\tau^2}{\partial^2\phi^*  \over \partial \eta^2}
- \nabla_{\bot}^{2}\phi^* +m^2 \phi^*=0~~.
\label{eq:E2.6}\end{eqnarray} 
These equations remain unchanged for the P-domain
and their first two terms change their sign in the domains L and R.
At $t^2-z^2>0$, these equations can be alternatively obtained as 
the equations of the Hamiltonian dynamic along the proper time $\tau$.
The canonical momenta conjugated to the fields $\phi$ and $\phi^*$ are    
\begin{eqnarray} 
\pi_{\phi}(x)={\delta(\sqrt{-{\rm g}}{\cal L})\over\delta \dot{\phi}(x)}
= \tau \dot{\phi}^*(x)
~~~ {\rm and}~~~
\pi_{\phi^*}(x)={\delta(\sqrt{-{\rm g}}{\cal L})\over\delta \dot{\phi}^*(x)}
= \tau \dot{\phi}^*(x)~,
\label{eq:E2.7}\end{eqnarray}  
 respectively.  The Hamiltonian of the field is as follows,
\begin{eqnarray} 
H=\int \tau d\eta d^2 {\bf r}\; [\tau^{-2}\pi_{\phi^*}\pi_{\phi}
+\tau^{-2} \partial_\eta \phi^* \partial_\eta\phi  
+\partial_x \phi^* \partial_x \phi +\partial_y \phi^* \partial_y \phi]~,  
\label{eq:E2.8}\end{eqnarray}  
 and the wave equations are just the Hamiltonian equations of
motion for the momenta.                                           

The explicit form of the one-particle wave functions for the scalar particles
and anti-particles is as follows,                            
\begin{eqnarray} 
\xi^{(\pm)}_{\nu,{\vec p}}(x)={e^{-\pi \nu /2}\over 2^{5/2}\pi }
H_{\mp i\nu}^{2 \choose 1} ( m_{\bot}\tau ) 
e^{\mp i\nu\eta} 
e^{\pm i{\vec p}{\vec r}}~.
\label{eq:E2.9}\end{eqnarray} 
where $ {\vec p}\equiv{\vec p}_\bot =(p_x,p_y)$,~ and 
$m_{\bot}^{2}= p_{\bot}^{2} +m^2$.        
In agreement with the  chosen dynamics,  the
eigenfunctions $~\xi_{\nu,p_\bot}(x)$  are  normalized on 
the space-like hypersurfaces $\tau=const$~ within
the future light wedge of the collision point:
\begin{eqnarray}
(\xi^{(\pm)*}_{\nu,p_\bot}, \xi^{(\pm)}_{\nu',p'_\bot})
=\int \tau~ d\eta ~d^2 {\vec r} ~ \xi^{(\pm)*}_{\nu,p_\bot}(x)
~i{\stackrel{\leftrightarrow}{\partial \over \partial \tau}} 
~\xi^{(\pm)}_{\nu',p'_\bot}(x)
=\delta (\nu -\nu')\delta ({\vec p}_{\bot}-{\vec p'}_{\bot}),\;\;\;\;
(\xi^{(\pm)*}_{\nu,p_\bot}, \xi^{(\mp)}_{\nu',p'_\bot})=0.
\label{eq:E2.10}\end{eqnarray} 
This norm is a consequence of the equation (\ref{eq:E2.5}) which expresses
the current conservation. To obtain it, one needs to integrate
(\ref{eq:E2.5}) over the 4-volume and transform the volume integral to the
integral over the closed surface. Thus we may need to continue the
solution found in the F-domain to the domains P, R and L. First, let us
notice that the wave packets,
\begin{eqnarray}
\Xi^{(\pm)}_{\theta,p_\bot}(x)=
{\mp i\over (2\pi)^{1/2}}\int_{-\infty}^{+\infty} d\nu
 e^{\mp i\nu \theta} \xi^{(\pm)}_{\nu,p_\bot}(x)
={1 \over 4\pi^{3/2}}  e^{\mp ip^0 t+ip^z z}
e^{\pm i{\vec p}_{\bot}{\vec r}_{\bot}}~, 
\label{eq:E2.11}
\end{eqnarray}           
can be easily recognized as the plane waves confined to within the F-domain.
The r.h.s. of the Eq.(\ref{eq:E2.11}) readily continue the solutions
to all $(t,z)$-plane. Performing the inverse transformation, one
obtains the following expressions,
\begin{eqnarray} 
\xi^{(\pm,F)}_{\nu,{\vec p}}(x)=\xi^{(\mp,P)}_{\nu,{\vec p}}(x)=
{e^{-\pi\nu/2}\over 2^{5/2}\pi }
H_{\mp i\nu}^{2\choose 1}
(m_{\bot}\tau) e^{\mp i\nu\eta} 
e^{\pm i{\vec p}{\vec r}},
\label{eq:E2.12}\end{eqnarray}      
\begin{eqnarray} 
\xi^{(\pm,L)}_{\nu,{\vec p}}(x)=\xi^{(\mp,R)}_{\nu,{\vec p}}(x)=
{\pm i e^{-\pi\nu/2}\over 2^{3/2} \pi^2}
K_{i\nu}(m_{\bot}\tau) e^{\mp i\nu\eta} 
e^{\pm i{\vec p}{\vec r}}~.
\label{eq:E2.13}\end{eqnarray}      
In what follows, I  shall denote this piecewise-defined function by one
symbol $\xi^{(\pm)}_{\nu,p_\bot}(x)$.  Using Eqs.(\ref{eq:E2.13})  one can
explicitly check that at the time-like hypersurfaces $\tau=const$ in the
domains L and R  the normal flux of the charge vanishes locally 
(since $~K_{i\nu}(m_{\bot}\tau)=K_{-i\nu}(m_{\bot}\tau)$ is a real 
function); the
states prepared via the initial data at the space-like hypersurface
$\tau=const$ in the P-domain  are predetermined to  penetrate the future
light cone through its vertex. The solutions of the relativistic field
equations are allowed to have discontinuities along the light cone
characteristics. Therefore, the problems with strong localization  of the
interaction domain in space and time, are of special kind. Placing the
vertex of the light cone (wedge) in the vertex of interaction, we can
discard any possible continuation of the partial waves out of the P- and
F-domains. 

The wave functions $~\xi^{(\pm)}_{\nu,p_\bot}(x)~$ given by 
Eqs.~(\ref{eq:E2.12}) and (\ref{eq:E2.13}) can be easily modified
in such a way that the coordinates of the wedge vertex will be an
additional parameter. Evidently, the system of these one particle solutions,
with the arbitrary coordinates of the vertex,
is isomorphic to the genuine system (\ref{eq:E2.11}) of the plane
waves $\Xi^{(\pm)}_{\theta,p_\bot}(x)$.  Therefore, with the vertex
coordinates explicitly retained, the extended set keeps the translational
symmetry inherent to the full set of the non-triggered events.

The functions $~\xi_{\nu,p_\bot}(x)$  can be employed for decomposition of 
the field:
\begin{equation} 
\phi (x)=\int d^2 {\vec r} d\nu [ a_{\nu,p_\bot}\xi^{(+)}_{\nu,p_\bot}(x)
+b^{\dag}_{\nu,p_\bot} \xi^{(-)}_{\nu,p_\bot}(x)]~,~~~
\phi^{\dag} (x)=\int d^2 {\vec r} d\nu 
[ a^{\dag}_{\nu,p_\bot}\xi^{(+)*}_{\nu,p_\bot}(x)
+b_{\nu,p_\bot} \xi^{(-)*}_{\nu,p_\bot}(x)]~,
\label{eq:E2.15}\end{equation}              
where we have passed to the symbolics which will become relevant after
the quantization.   This symbolics implies that we have already gave the
definitions of the particle and anti-particle states. 

The plane waves (\ref{eq:E2.11}) can be equally
used to  decompose the field operators,
\begin{equation} 
\phi (x)=\int d^2 {\vec p_\bot} d \theta 
[ \alpha_{\theta,p_\bot}\Xi^{(+)}_{\theta,p_\bot}(x)
+\beta^{\dag}_{\theta,p_\bot} \Xi^{(-)}_{\theta,p_\bot}(x)], \;\;\;
\phi^{\dag} (x)=\int d^2 {\vec p_\bot} d \theta 
[\alpha^{\dag}_{\theta,p_\bot}\Xi^{(+)*}_{\theta, p_\bot}(x)
+\beta_{\theta,p_\bot} \Xi^{(-)*}_{\theta,p_\bot}(x)],
\label{eq:E2.16}\end{equation}        
thus providing the common language of the momentum decomposition.          
In our case this language is somewhat restricted because in the geometry
of localized interaction we miss the quantum operator of the conserved
momentum.

However, mathematically,  we deal with the solutions of the homogeneous
wave equations in  free space. Thus the correlators remain unchanged.
For example, in virtue of $d\theta = dp_z/p^0$, we have
\begin{eqnarray}
D_{10}(x,y)= -i\langle 0|\phi(x)\phi^*(y)|0\rangle=
- i\int d\theta d^2 {\vec p}_{\bot} \Xi^{(+)}_{y,p_\bot}(x)
\Xi^{(+)*}_{y,p_\bot}(y)~,\nonumber \\
D_{01}(x,y)= -i\langle 0|\phi^*(y)\phi(x)|0\rangle=
- i\int d\theta d^2 {\vec p}_{\bot} \Xi^{(-)}_{y,p_\bot}(x)
\Xi^{(-)*}_{y,p_\bot}(y)~. 
\label{eq:E2.17}\end{eqnarray} 
In the momentum representation we have,  
\begin{eqnarray} 
D_{\stackrel{\scriptscriptstyle 10}{\scriptscriptstyle 01}}(x,y)
=\int {d^4 p \over (2\pi)^4} e^{-ip(x-y)} 
[ -2\pi i\delta (p^2-m^2)\theta(\pm p^0)]~.      
\label{eq:E2.17a}\end{eqnarray}    
One may easily recognize the standard expression for  $D_{10}(p)$ and
$D_{01}(p)$ in the integrand. The latter represents the density of states
for the final-state particles and anti-particles in the initially
unoccupied vacuum.  Dependence of  $D_{10}(x,x')$ and $D_{01}(x,x')$
only on the difference $~x-x'~$ is fully consistent with the free
field dynamics.

These two Wightman functions lead to a familiar form of various
propagators. Introducing the commutator $D_{0}=D_{10}-D_{01}$,  we have 
in coordinate form:
\begin{eqnarray}
D_{\stackrel{\scriptscriptstyle ret}{\scriptscriptstyle adv}}(x,x')
=\theta[\pm (x_0-x'_0)] D_0(x-x')~,\nonumber \\      
D_{00}(x,x')=-i\langle 0|T\phi(x)\phi^*(y)|0\rangle=
\theta(x_0-x'_0)D_{10}(x,x')+\theta(x'_0-x_0)D_{01}(x,x')~,  \nonumber \\
D_{11}(x,x')=-i\langle 0|T^{\dag}\phi(x)\phi^*(y)|0\rangle =
\theta(x'_0-x_0)D_{10}(x,x')+
\theta(x_0-x'_0)D_{01}(x,x')~, 
\label{eq:E2.17b}\end{eqnarray}  
so that in momentum representation we have
\begin{eqnarray} 
D_{\stackrel{\scriptscriptstyle ret}{\scriptscriptstyle adv}}(p) =
{1\over (p_0\pm i0)^2-{\bf p}^2 -m^2}~,~~~
D_{\stackrel{\scriptscriptstyle 00}{\scriptscriptstyle 11}}(p) =
{\pm 1\over p_{0}^{2}-{\bf p}^2 -m^2 \pm i0}~.
\label{eq:E2.17c}\end{eqnarray} 
Since the commutator $D_0(x,x')$ vanishes outside the light cone, the 
theta-functions in the definition of the retarded and advanced
propagators can be conveniently rewritten as $\theta[\pm (\tau-\tau')]$.
In virtue of $D_{00}=D_{ret}+D_{01}$ and  $D_{11}=D_{10}-D_{ret}$, the
other theta-functions acquire the same invariant definitions.
\subsection{Quantization}
\label{subsec:SB22}   

The canonical commutation relations between the field coordinates and momenta
$\phi$ and $\pi_{\phi}$ ,$\phi^*$ and $\pi_{\phi^*}$, read as
\begin{eqnarray}
\tau[\phi(\eta,\vec{r}),\dot{\phi}^*(\eta',\vec{r'}]=
\delta (\eta-\eta') \delta(\vec{r}-\vec{r'}); \nonumber  \\
\tau[\phi^*(\eta,\vec{r}),\dot{\phi}(\eta',\vec{r'}]=
\delta (\eta-\eta') \delta(\vec{r}-\vec{r'}).
\label{eq:E2.18}\end{eqnarray}  
They are satisfied if and only if the commutation relations between the Fock
operators are of the standard form,      
\begin{eqnarray}
[a_{\nu,p_\bot},a^{\dag}_{\nu',\vec{p}'}]=
[b_{\nu,p_\bot},b^{\dag}_{\nu',\vec{p}'}]=\delta(\nu-\nu')
\delta(\vec{p}-\vec{p}'), 
\label{eq:E2.19}\end{eqnarray}       
and
\begin{eqnarray}  
[\alpha_{\theta,\vec{p}},\alpha^{\dag}_{\theta',\vec{p}'}]=
[\beta_{\theta,\vec{p}},\beta^{\dag}_{\theta',\vec{p}'}]=\delta(\theta-\theta')
\delta(\vec{p}-\vec{p}')
\label{eq:E2.20}\end{eqnarray}        

\section{States of Fermions}
\label{sec:SN4}  

The states of fermions are studied along the same lines as the states of
the bosons. A minor complication is due to the nontrivial form of the
covariant derivative of the spinor in the curvilinear coordinates.
The spinor field is essentially defined in the tangent space and,
therefore, its covariant derivative should be calculated in 
the so-called tetrad formalism \cite{Fock,Witten}.

The covariant derivative of the tetrad vector
includes two connections (gauge fields). One of them, 
$\Gamma^{\lambda}_{~\mu\nu}(x)$, is the gauge  field which provides 
covariance with respect to the general transformation of
coordinates. The second gauge field, $\omega_{\mu}^{~ab}(x)$, the
spin connection, provides covariance with respect to the local
Lorentz rotation. 
                                          
Let $x^\mu =(\tau,x,y,\eta)$ be the contravariant components of the
curvilinear coordinates and $x^a =(t,x,y,z)\equiv(x^0,x^1,x^2,x^3)$ those
of the flat Minkowsky space. Then the tetrad vectors $e^{a}_{~\mu}$ can be
taken as follows,
\begin{eqnarray}  
e^{0}_{~\mu}=(1,0,0,0),~~e^{1}_{~\mu}=(0,1,0,0),~~
e^{2}_{~\mu}=(0,0,1,0),~~e^{3}_{~\mu}=(0,0,0,\tau)~.
\label{eq:E3.1}\end{eqnarray}        
They correctly reproduce the curvilinear metric ${\rm g}_{\mu\nu}$ and 
the flat Minkowsky metric $g_{ab}$:
\begin{eqnarray}
{\rm g}_{\mu\nu}=g_{ab}e^{a}_{~\mu}e^{b}_{~\nu}
={\rm diag}[1,-1,-1,-\tau^2]~,~~~
g^{ab}= {\rm g}^{\mu\nu} e^{a}_{~\mu}e^{b}_{~\nu}={\rm diag}[1,-1,-1,-1] ~.
\label{eq:E3.2}\end{eqnarray} 
The spin connection can be found from the condition that the covariant
derivatives of the tetrad vectors equal to zero,
\begin{eqnarray}
\nabla_\mu e^{a}_{~\nu}=\partial_\mu e^{a}_{~\nu}
+\omega^{~a}_{\mu~b}e^{b}_{~\nu} -
\Gamma^{\lambda}_{~\mu\nu}e^{a}_{~\lambda} =0~~,\label{eq:E3.3} \\
\Gamma^{\lambda}_{~\mu\nu}={1\over 2} {\rm g}^{\lambda\rho}
\bigg[{\partial {\rm g}_{\rho\mu} \over \partial x^\nu }
+{\partial {\rm g}_{\rho\nu} \over \partial x^\mu }
-{\partial {\rm g}_{\mu\nu} \over \partial x^\rho } \bigg]~~.\nonumber 
\end{eqnarray} 
The covariant derivative of the spinor field includes only the spin
connection,
 \begin{eqnarray}  
\nabla_\mu\psi(x)= \big[\partial_\mu +
{1\over 4}\omega_{\mu}^{~ab}(x)\Sigma_{ab}\big]\psi(x)~,
\label{eq:E3.4}\end{eqnarray} 
where $\Sigma^{ab}={1\over 2}[\gamma^a\gamma^b-\gamma^b\gamma^a]$ is
an obvious generator of the Lorentz rotations and $\gamma^a$ are the
Dirac matrices of Minkowsky space. Introducing the Dirac matrices
in curvilinear coordinates, $\gamma^{\mu}(x)=e^{\mu}_{~a}(x)\gamma^a$, 
one obtains the Dirac equation in curvilinear coordinates,       
\begin{eqnarray}  
[\gamma^\mu(x)(i\nabla_\mu +gA_\mu(x))-m]\psi(x)=0 ~,
\label{eq:E3.5}\end{eqnarray}   
where $A^\mu(x)$ is the gauge field associated with the local group of 
the internal
symmetry. The conjugated spinor is defined as usually,
${\overline \psi}=\psi^{\dag}\gamma^0$, and obeys the equation, 
\begin{eqnarray}  
(-i\nabla_\mu +gA_\mu(x)){\overline\psi}(x)\gamma^\mu(x)- 
m{\overline\psi}(x)=0 ~~.
\label{eq:E3.5a}\end{eqnarray}  
 
These two Dirac equations correspond to the action, 
\begin{eqnarray}
{\cal A}=\int d^4 x \sqrt{-g}{\cal L}(x)=
\int d^4 x \sqrt{-g} \{ {i\over 2}[{\overline\psi}\gamma^\mu(x)
\nabla_\mu\psi- (\nabla_\mu {\overline\psi})\gamma^\mu(x)\psi]
+ g{\overline\psi}\gamma^\mu(x) A_\mu\psi -
m{\overline\psi}\psi \}~~,
\label{eq:E3.5b}    
\end{eqnarray} 
from which one  easily obtains the locally conserved $U(1)$-current,         
\begin{eqnarray}
J^\mu(x)= {\overline\psi}(x)\gamma^\mu(x)\psi(x)~, 
~~~~ (-{\rm g})^{-1/2}\partial_\mu [(-{\rm g})^{1/2}
{\rm g}^{\mu\nu}(x)J_\nu (x)]=0~.
\label{eq:E3.5c}\end{eqnarray}  
The Dirac equations (\ref{eq:E3.5}) and (\ref{eq:E3.5a}) can be 
alternatively obtained as the equations of the
Hamiltonian dynamics along the proper time $\tau$.
The canonical momenta conjugated to the fields $\psi$ and ${\overline\psi}$
are                                 
\begin{eqnarray} 
\pi_{\psi}(x)={\delta(\sqrt{-{\rm g}}{\cal L})\over\delta \dot{\psi}(x)}
= {i\tau\over 2} {\overline\psi}(x)\gamma^0
\;\; {\rm and} \;\;\;
\pi_{{\overline\psi}}(x)=
{\delta(\sqrt{-{\rm g}}{\cal L})\over\delta \dot{{\overline\psi}(x)} }
= -{i\tau\over 2} \gamma^0\psi(x)~,
\label{eq:E3.5d}\end{eqnarray}  
 respectively.
The Hamiltonian of the Dirac field in the wedge dynamics has the following
form,
\begin{eqnarray} 
H=\int \tau d\eta d^2 {\vec r}\;\sqrt{-{\rm g}} 
\{ -{i\over 2}[{\overline\psi}\gamma^i(x)
\nabla_i\psi- (\nabla_i {\overline\psi})\gamma^i (x)\psi]
- g{\overline\psi}\gamma^\mu(x) A_\mu\psi+
m{\overline\psi}\psi \}~~.
\label{eq:E3.5e}\end{eqnarray}  
 and the wave equations are just the Hamiltonian equations of
motion for the momenta.                                           
 
The non-vanishing components of the connections are 
$\Gamma^{\cdot}_{\eta\tau\eta} = \Gamma^{\cdot}_{\eta\eta\tau}=
-\Gamma^{\cdot}_{\tau\eta\eta}=-\tau$~ and ~$\omega_{\eta}^{~30}
=-\omega_{\eta}^{~03}=1$. Moreover, we have $\gamma^\tau(x)=\gamma^0$
and  $\gamma^\eta(x)=\tau^{-1}\gamma^3 $. The explicit form of the
Dirac equation in our case is as follows,
\begin{eqnarray}  
[i \not\!\nabla -m]\psi(x)= [i\gamma^0(\partial_\tau +{1\over 2\tau})
+i\gamma^3{1\over \tau}\partial_\eta +i \gamma^r \partial_r -m]\psi(x)=0~~.
\label{eq:E3.6}\end{eqnarray}     
The solutions to this equation will be looked in the form 
$\psi(x)= [i \not\!\nabla +m]\chi(x)$, with the function $\chi(x)$ that
obeys the ``squared'' Dirac equation,
\begin{eqnarray}  
[i \not\!\nabla +m][i \not\!\nabla -m]\chi(x)= 
\bigg[\partial_{\tau}^{2} +{1\over \tau}\partial_\tau
-{1\over \tau^2}\partial_{\eta}^{2} - \partial_{r}^{2} +m^2-
\gamma^0\gamma^3 {1\over \tau^2}\partial_{\eta} \bigg]\chi(x)=0~~.
\label{eq:E3.7}\end{eqnarray}       
The spinor part $\beta_\sigma$ of the function $\chi(x)$ can be chosen
as  an eigen-function of the operator $\gamma^0\gamma^3 $, {\em viz.},
$~\gamma^0\gamma^3\beta_{\sigma} = \beta_{\sigma}~,~~\sigma=1,2~$.         
Therefore, the solution of the original Dirac equation can be written down 
as $~\psi^{\pm}_{\sigma} = w_\sigma \chi^{\pm}(x)$,~   
with the bi-spinor operators $w_\sigma=[i \not\!\nabla +m]\beta_\sigma $
that act on the positive- and negative-frequency solutions $\chi^{\pm}(x)$
of the scalar equation
\begin{eqnarray}  
\bigg[\partial_{\tau}^{2} +{1\over \tau}\partial_\tau
-{1\over \tau^2}(\partial_{\eta}+{1\over 2})^{2} - \partial_{r}^{2} 
+m^2\bigg]\chi^{\pm}(x)=0~~.
\label{eq:E3.10}\end{eqnarray}       
 Using the spinor representation of the gamma-matrices,
the explicit form of the spinors is found as follows,
\begin{eqnarray}  
w_{1}({\vec p},{\hat l})=   \left( \begin{array}{c} 
                             m \\ 
                             0 \\ 
                         i{\hat l} \\
                          i\partial_{+}
                             \end{array} \right);  \;\;\;\;  
w_{2}({\vec p},{\hat l}) =   \left( \begin{array}{c}  
                          -i\partial_{-} \\ 
                           i{\hat l} \\
                              0 \\
                               m \\
                             \end{array} \right)~~,
\label{eq:E3.11}\end{eqnarray}          
where the  operators,
\begin{eqnarray}  
{\hat l}=\partial_\tau + {1\over \tau}(\partial_\eta +{1\over 2})~~~
{\rm and}~~~ \partial_{\pm}=\partial_x \pm i\partial_y~~,
\label{eq:E3.12}\end{eqnarray}  
is convenient to preserve in the differential form.  
The full set of the one-particle solution can be conveniently written down
as follows,
\begin{eqnarray}  
\omega^{(\pm)}_{\sigma,\nu,{\vec p}}(x)= 
{  e^{-\pi\nu/2} 
 w_{\sigma}({\vec p},\nu)  \over
2^{5/2} \pi m_{\bot}^{1/2} } e^{\pm i\nu \eta } e^{\pm i{\vec p}{\vec r}}
H_{\mp i\nu -1/2}^{2\choose 1}(m_{\bot}\tau)~~,
\label{eq:E3.13} \end{eqnarray} 
The spinors $w_{\sigma}({\vec p},\nu)$  are those of Eq.(\ref{eq:E3.11})
with the replacement of  spacial derivatives by  corresponding
momenta.
These solutions are normalized according to the current conservation law.
This leads to the scalar product of the following form,
\begin{eqnarray}  
(\psi_1,\psi_2)=
\int \tau d\eta d^2 {\vec r} 
{\overline \psi}_1 (\tau,\eta,{\vec r})\gamma^\tau  
\psi_2 (\tau,\eta,{\vec r})~~.
\label{eq:E3.14}\end{eqnarray}   
With this definition, the Dirac equation is self-adjoint.
The orthonormality relations are read as follows,
\begin{eqnarray}  
(\omega^{(\pm)}_{\sigma,\nu,{\vec p}}, 
\omega^{(\pm)}_{\sigma',\nu',{\vec p}'})= \delta_{\sigma \sigma'}
\delta({\vec p}-{\vec p}')\delta(\nu-\nu')~,~~~
 (\omega^{(\pm)}_{\sigma,\nu,{\vec p}}, 
\omega^{(\mp)}_{\sigma',\nu',{\vec p}'})= 0~~.
\label{eq:E3.15}\end{eqnarray}      

An equivalent set of the one-particle solutions is obtained by means of
the Fourier transform,
\begin{eqnarray}  
\Omega^{(\pm)}_{\sigma,\theta,{\vec p}}(x)= \pm e^{\mp i\pi/4}
\int_{-\infty}^{+\infty}  {d\theta \over (2\pi)^{1/2}} e^{\mp i\nu\theta}
\omega^{(\pm)}_{\sigma, \nu, {\vec p}}(x)~.
\label{eq:E3.16} \end{eqnarray}  
Instead of the boost quantum number $\nu$, these solutions carry the
rapidity quantum number $\theta$, which parameterize the energy and the
longitudinal momentum of the on-mass-shell fermion as
$p^0=m_{\bot}\cosh\theta$ and $p^3=m_{\bot}\sinh\theta$, respectively.  
In this way we obtain           
the positive- and negative-frequency plane wave solutions,
\begin{eqnarray}  
\Omega^{(\pm)}_{\sigma,\theta,{\vec p}}=
  { w^{(\pm)}_{\sigma}({\vec p},{\hat l})e^{(\theta-\eta) /2}
  \over 2^{1/2}(2\pi)^{3/2} m_{\bot}^{1/2} } 
  e^{\mp im_{\bot}\tau\cosh(\theta-\eta)}
  e^{\pm i{\vec p}{\vec r}}~,   
\label{eq:E3.17}\end{eqnarray}    
which are confined to within the future light wedge of the
collision plane $t=z=0$. They are normalized and orthogonal according
to relations, 
\begin{eqnarray}  
(\Omega^{(\pm)}_{\sigma,\theta,{\vec p}}, 
\Omega^{(\pm)}_{\sigma',\theta',{\vec p}'})= \delta_{\sigma \sigma'}
\delta({\vec p}-{\vec p}')\delta(\theta-\theta')~,~~~
 (\Omega^{(\pm)}_{\sigma,\theta,{\vec p}}, 
\Omega^{(\mp)}_{\sigma',\theta',{\vec p}'})= 0~~.
\label{eq:E3.18}\end{eqnarray} 
Action of the operator $~{\hat l}~$ onto the function $e^{-\eta/2}f(\eta)$
is as follows: 
\begin{eqnarray} 
{\hat l}e^{-\eta/2}f(\eta)=e^{-\eta/2}(\partial_\tau+
\tau^{-1}\partial_\eta)=e^{\eta/2}(\partial_t+\partial_z)~,\nonumber  
\end{eqnarray}       
and the spinors preserve their original form up to the change of variables in
operator ${\hat l}$. Eventually, the solutions (\ref{eq:E3.17}) can
be rewritten in the form,
\begin{eqnarray}  
\Omega^{(\pm)}_{\sigma,\theta,{\vec p}}(x)= 
\Lambda (-\eta)\psi^{(\pm)}_{\sigma{\bf p}}(x)~~,
\label{eq:E3.19}\end{eqnarray}      
where
\begin{eqnarray} 
\Lambda (\eta)=\cosh(\eta/2) + \gamma^0\gamma^3\sinh(\eta/2)=
{\rm diag}[e^{\eta/2},e^{-\eta/2},e^{-\eta/2},e^{\eta/2}]~.\nonumber  
\end{eqnarray}   
is the matrix of the spinor Lorentz rotation with rapidity $\eta$,
and $\psi^{(\pm)}_{\sigma{\bf p}}$ are the standard plane wave solutions
of the Dirac equation normalized, however, on the hypersurfaces
$\tau=const$:
\begin{eqnarray}  
\psi^{(\pm)}_{\sigma{\bf p}}(x)=
{ u^{(\pm)}_{\sigma}({\vec p},\theta)e^{\theta /2}
  \over 2^{1/2}(2\pi)^{3/2} m_{\bot}^{1/2} } 
e^{\mp im_{\bot}\tau\cosh(\theta-\eta)} e^{\pm i{\vec p}{\vec r}}~,  
\label{eq:E3.20} \\ 
u_{1}^{(+)}({\vec p},\theta)=   
                        \left( \begin{array}{c} 
                             m \\ 
                             0 \\ 
                            m_{\bot} e^{-\theta /2}\\
                           -(p_x+ip_y)
                             \end{array} \right)~, ~~~  
u_{2}^{(+)}({\vec p},\theta) =   \left( \begin{array}{c}  
                           p_x-ip_y \\ 
                           m_{\bot} e^{-\theta /2} \\
                              0 \\
                               m \\
                             \end{array} \right)~, \nonumber \\  
u_{1}^{(-)}({\vec p},\theta)=   
                        \left( \begin{array}{c} 
                             m \\ 
                             0 \\ 
                           - m_{\bot} e^{-\theta /2}\\
                             p_x+ip_y
                             \end{array} \right)~, ~~~  
u_{2}^{(-)}({\vec p},\theta) =   \left( \begin{array}{c}  
                           -(p_x-ip_y) \\ 
                           -m_{\bot} e^{-\theta /2} \\
                              0 \\
                               m \\
                             \end{array} \right)~. \nonumber   
\end{eqnarray}          
Either set of the one-particle solutions can be used in order to compute
various correlators of the free spinor field. The result reads as follows. 
The two Wightman correlators,
\begin{eqnarray} 
 G_{10}(x_1,x_2)= -i\langle 0|\psi(x_1)\overline{\psi}(x_2)|0\rangle =
 - i\int d\theta d^2 {\vec p} \Omega^{(+)}_{\theta,{\vec p}}(x_1)
\overline{\Omega}^{(+)}_{\theta,{\vec p}}(x_2)~, \nonumber        
\end{eqnarray}  
and
\begin{eqnarray} 
G_{01}(x_1,x_2)= i\langle 0|\overline{\psi}(x_2)\psi(x_1)|0\rangle=
 i\int d\theta d^2 {\vec p} \Omega^{(-)}_{\theta,{\vec p}}(x_1)
\overline{\Omega}^{(-)}_{\theta,{\vec p}}(x_2)~, \nonumber    
\end{eqnarray}  
have the Fourier representation,
\begin{eqnarray}  
G_{\stackrel{\scriptscriptstyle 10}{\scriptscriptstyle 01}}(x_1,x_2)
=\int {d^4 p \over (2\pi)^4} e^{-ip(x-x')} 
[ -2\pi i\delta (p^2-m^2)\theta(\pm p^0)]
\Lambda(-\eta_1)(\not\! p +m)\Lambda(\eta_2)~.   
\label{eq:E3.21}\end{eqnarray}      
All other correlators and propagators are easily found via these two
following the guideline of the previous section.

Presence of the matrices $\Lambda$ which perform the local Lorentz
rotation of the spinors is vital for consistency between the gauge
transformations of the spinors and of their gauge field. Local Lorentz 
rotation (\ref{eq:E3.19}) of the spinors is in one-to-one correspondence
with the ``rotation'' of the gauge condition $A^\tau (x)=0$. As a result,
the matrices  $\Lambda(\eta)$ from the fermion correlators  can be
absorbed into the vertex of interaction. Indeed, for the closed fermion
loop, every vertex of
interaction between the fermion and the gauge field appears only between
the two fermion correlators. In the same way, every fermion line connects
two vertices. Therefore, the proof is as follows,
\begin{eqnarray}  
\Lambda(\eta)(\gamma^\tau A_\tau +\gamma^\eta A_\eta)\Lambda(-\eta)=
\gamma^0 \big(\cosh\eta A_\tau - {\sinh\eta\over\tau} A_\eta\big)
+ \gamma^3 \big(-\sinh\eta A_\tau + {\cosh\eta\over\tau} A_\eta\big)
=\gamma^0 A_0 +\gamma^3 A_3~,   
\label{eq:E3.23}\end{eqnarray}        
and it results in the standard form of the covariant interaction vertex. 
The latter is  affected only by a specific choice of the gauge for the field
$A^\mu(x)$.  In the composite spinor operators, like fermion self-energy,
the matrices $\Lambda$ of local rotation should be retained explicitly.

\section{Summary}

I suggest a new approach to the study the dynamics of the high-energy
deeply inelastic processes with the aim to overcome limitations imposed by
the parton model. The new theory  deals with a single
Hilbert space  for all initial-  and final-state particles. The method is
expected to work even when no scale of the hard probe is  specified  and
the standard factorization scheme is inapplicable. The approach is
designed exclusively for collisions at extreme energies and requires
special selection of events, those where the short-scale structure of the
colliding systems is resolved and spectrum of the secondaries has a
rapidity plateau. I emphasize the role of a trigger in
collection the data  and show that special selection of events may
significantly affect the type of the available dynamical information.

\vskip 3cm
\centerline {\bf ACKNOWLEDGEMENTS}

\bigskip 

I am grateful to D. Dyakonov, B. Muller, L. McLerran, E.Shuryak,  
E.Surdutovich and R.Venugopalan for many stimulating discussions. I much
profitted  from intensive discussions during the International Workshop on
Quantum Chromodynamics and Ultrarelativistic Heavy Ion Collisions at the
ECT${^*}$ in Trento, Italy.

This work was supported by the U.S. Department of Energy under Contract 
No. DE--FG02--94ER40831.

\end{document}